\documentclass{article}
\usepackage{amsmath,amsthm,amsfonts,amscd}
\DeclareMathOperator{\Hom}{Hom}
\DeclareMathOperator{\Com}{Com}
\DeclareMathOperator{\II}{I}
\DeclareMathOperator{\1}{id}
\DeclareMathOperator{\ad}{ad}
\DeclareMathOperator{\Ker}{Ker}

\DeclareMathOperator{\ch}{char}
\newcommand{\NN}{\mathbb{N}}
\newcommand{\EEnd}{\mathcal Com}
\newcommand{\EE}{\mathcal C}
\newcommand{\bul}{\bullet}

\newcommand{\w}{\omega}
\newcommand{\p}{\partial}
\renewcommand{\=}{:=}
\renewcommand{\t}{\otimes}
\renewcommand{\o}{\circ}

\newtheorem{thm}{Theorem}
\theoremstyle{definition}
 \newtheorem{defn}[thm]{Definition}
\theoremstyle{definition}
 \newtheorem{exam}[thm]{Example}
\begin{document}
\title{Operadic curvature as a tool for gravity}
\author{Eugen Paal\\ \\
Department of Mathematics, Tallinn Technical University\\
Ehitajate tee 5, 19086 Tallinn, Estonia\\
e-mail: eugen@edu.ttu.ee
}
\date{}
%
\maketitle
\thispagestyle{empty}
\begin{abstract}
The deformation equation and its integrability condition (Bianchi identity)
of a non-associative deformation in operad algebra are found. Their relation
to the theory of gravity is discussed.
\end{abstract}

\section{Introduction}

Non-associativity is sometimes said to be an \emph{algebraic} equivalent of
the differential geometrical concept of curvature (e.~g.
\cite{NeSa1,NeSa2}). By adjusting this for physics, one may surmise that
gravity and gauge fields geometry can be described in algebraic terms. In
particular, instead of the \emph{curvature} of the space-time,
\emph{associator} rises to the fore \cite{Ki,Ak}. In this sense, gravity can
be seen to have an algebraic  representation. When non-associativity of
space-time becomes large, operadic structure will become important and one
must use \emph{operad algebra} to understand the algebraic underground of
the gravity and how the gravity could be quantized. Instead of the
\emph{quantum} gravity, the \emph{operadic gravity} rises to the fore.

In this paper, the equivalence is clarified from the \emph{linear} deformation
theoretical point of view. By using the Gerstenhaber brackets and a coboundary
operator in a pre-operad, the (formal) associator can be represented
as a curvature form in differential geometry. This (structure) equation is called a
\emph{deformation equation}.
Its integrability condition is the Bianchi identity. Their
relation to the theory of gravity is discussed.

\section{Operad algebra}

Let $K$ be a unital associative commutative ring, $\ch K\neq2,3$, and let $C^n$
($n\in\NN$) be unital $K$-modules. For \emph{homogeneous} $f\in C^n$,
$n$ is called the \emph{degree} of $f$ and (when it does not
cause confusion) $f$ is written instead of $\deg f$. For example, $(-1)^f\=(-1)^n$,
$C^f\=C^n$ and $\o_f\=\o_n$. Also, it is convenient to use the
\emph{reduced} degree $|f|\=n-1$. Throughout this paper, it is assumed that
$\t\=\t_K$.

\begin{defn}
A linear \emph{pre-operad} (\emph{composition system}) with coefficients in
$K$ is a sequence $C\=\{C^n\}_{n\in\NN}$ of unital $K$-modules (an
$\NN$-graded $K$-module), such that the following conditions hold.
\begin{enumerate}
\item[(1)]
For $0\leq i\leq m-1$ there exist \emph{(partial) compositions}
\[
  \o_i\in\Hom(C^m\t C^n,C^{m+n-1}),\qquad |\o_i|=0.
\]
\item[(2)]
For all $h\t f\t g\in C^h\t C^f\t C^g$, the \emph{composition
(associativity) relations} hold,
\[
(h\o_i f)\o_j g=
\begin{cases}
    (-1)^{|f||g|} (h\o_j g)\o_{i+|g|}f
                       &\text{if $0\leq j\leq i-1$},\\
    h\o_i(f\o_{j-i}g)  &\text{if $i\leq j\leq i+|f|$},\\
    (-1)^{|f||g|}(h\o_{j-|f|}g)\o_i f
                       &\text{if $i+f\leq j\leq|h|+|f|$}.
\end{cases}
\]
\item[(3)]
There exists a unit $\II\in C^1$ such that
\[
\II\o_0 f=f=f\o_i \II,\qquad 0\leq i\leq |f|.
\]
\end{enumerate}
\end{defn}

In the 2nd item, the \emph{first} and \emph{third} parts of the defining
relations turn out to be equivalent.

Elements of an operad may be called \emph{operations}.
Operad can be seen as a system of operations closed with respect to
compositions.

\begin{exam}[composition pre-operad \rm{\cite{Ger}}]
\label{CG} Let $L$ be a unital $K$-module and
$\EE_L^n\={\EEnd}_L^n\=\Hom(L^{\t n},L)$. Define the partial compositions
for $f\t g\in\EE_L^f\t\EE_L^g$ as
\[
f\o_i g\=(-1)^{i|g|}f\o(\1_L^{\t i}\t g\t\1_L^{\t(|f|-i)}),
         \qquad 0\leq i\leq|f|.
\]
Then $\EE_L\=\{\EE_L^n\}_{n\in\NN}$ is a pre-operad (with the unit
$\1_L\in\EE_L^1$) called the \emph{composition pre-operad} of $L$.
\end{exam}

\section{Gerstenhaber brackets}

The \emph{total composition} $\bul\:C^f\t C^g\to C^{f+|g|}$ is defined by
\[
f\bul g\=\sum_{i=0}^{|f|}f\o_i g\in C^{f+|g|},
     \qquad |\bul|=0.
\]
The pair $\Com C\=\{C,\bul\}$ is called the \emph{composition algebra} of $C$.

The \emph{Gerstenhaber brackets} $[\cdot,\cdot]$ are defined in $\Com C$ by
\[
[f,g]\=f\bul g-(-1)^{|f||g|}g\bul f=-(-1)^{|f||g|}[g,f],\qquad|[\cdot,\cdot]|=0.
\]
The \emph{commutator algebra} of $\Com C$ is denoted as
$\Com^{-}\!C\=\{C,[\cdot,\cdot]\}$. It turns out that $\Com^-\!C$ is a
\emph{graded Lie algebra}. The Jacobi
identity reads
\[
(-1)^{|f||h|}[[f,g],h]+(-1)^{|g||f|}[[g,h],f]+(-1)^{|h||g|}[[h,f],g]=0.
\]

In a pre-operad $C$, define a \emph{pre-coboundary} operator $\p_\Delta$ by
\begin{align*}
\p_\Delta f&\=\ad_\Delta^{right}f\=[f,\Delta],\qquad |\p_\Delta|=|\Delta|.
\end{align*}
It follows from the Jacobi identity the (right) derivation property
\[
\p_\Delta[f,g]=(-1)^{|\Delta||g|}[\p_\Delta f,g]+[f,\p_\Delta g]
\]
and the commutation relation
\[
[\p_f,\p_g]:=\p_f\p_g-(-1)^{|f||g|}\p_g\p_f=\p_{[g,f]}.
\]
Thus, if $|\Delta|$ is \emph{odd}, then
\[
\p_\Delta^{2}=\frac{1}{2}[\p_\Delta,\p_\Delta]=
\frac{1}{2}\p_{[\Delta,\Delta]}=\p_{\Delta\bul\Delta}=\p_{\Delta^{2}}.
\]

\section{Deformation equation}

For an operad $C$, let $\Delta,\Delta_0\in C^{2}$.
The difference $\w:=\Delta-\Delta_0$
is called a \emph{deformation}, and $\Delta$ is said to be a deformation of
$\Delta_0$. Let $\p:=\p_{\Delta_0}$, and denote the (formal) associators of
$\Delta$ and $\Delta_0$ as follows:
\[
A:=\Delta\bul\Delta=\frac{1}{2}[\Delta,\Delta],
\qquad
A_0:=\Delta_0\bul\Delta_0=\frac{1}{2}[\Delta_0,\Delta_0].
\]
The deformation is called \emph{associative} if $A=0=A_0$.

To find the deformation equation, calculate
\begin{align*}
A
&=\frac{1}{2}[\Delta_0+\w,\Delta_0+\w]\\
&=\frac{1}{2}[\Delta_0,\Delta_0]+\frac{1}{2}[\Delta_0,\w]
     +\frac{1}{2}[\w,\Delta_0]+\frac{1}{2}[\w,\w]\\
&=A_0-\frac{1}{2}(-1)^{|\Delta_0||\w|}[\w,\Delta_0]
     +\frac{1}{2}[\w,\Delta_0]+\frac{1}{2}[\w,\w]\\
&=A_0+[\w,\Delta_0]+\frac{1}{2}[\w,\w].
\end{align*}
So we get the \emph{deformation equation}
$$
\boxed{A-A_0=\p\w+\frac{1}{2}[\w,\w]}
$$
The deformation equation can be seen as a differential equation for
$\omega$ with given associators $A_0, A$. Note that if the associator is
fixed, i.~e. $A=A_0$, one obtains the \emph{Maurer-Cartan (master)
equations}, well-known from the theory of \emph{associative} deformations.

\section{Prolongation}

Now differentiate the deformation equation,
\begin{align*}
\p(A-A_0)
&=\p^{2}\w+\frac{1}{2}\p[\w,\w]\\
&=\p^{2}\w+\frac{1}{2}(-1)^{|\p||\w|}[\p\w,\w]
     +\frac{1}{2}[\w,\p\w]\\
&=\p^{2}\w-\frac{1}{2}[\p\w,\w]
     +\frac{1}{2}[\w,\p\w]\\
&=\p^{2}\w-\frac{1}{2}[\p\w,\w]
     -\frac{1}{2}(-1)^{|\p\w||\w|}[\p\w,\w]\\
&=\p^{2}\w-[\p\w,\w].
\end{align*}
Again using the deformation equation, we obtain
\begin{align*}
\p(A-A_0)
&=\p^{2}\w-[\p\w,\w]\\
&=\p^{2}\w-[A-A_0-\frac{1}{2}[\w,\w],\w]\\
&=\p^{2}\w-[A-A_0,\w]+\frac{1}{2}[[\w,\w],\w].
\end{align*}
Finally use $[[\w,\w],\w]=0$ to obtain the condition
$$
\boxed{\p(A-A_0)=\p^{2}\w-[A-A_0,\w] }
$$

\section{Associativity constraint and Bianchi identity}

We know that $\p^{2}=\p_{A_0}$. Hence, if \emph{associativity} constraint $A_0=0$ holds,
then
\[
\boxed{\p^{2}=0}
\]
The deformation equation for such a \emph{non-associative} deformation
reads
\[
\boxed{A=\p\w+\frac{1}{2}[\w,\w] }
\]
One can see that associator is a formal \emph{curvature} while the
deformation is working as a \emph{connection}. One can say that associator
is an \emph{operadic} equivalent of the curvature.
The integrability condition of the deformation equation reads as the
\emph{Bianchi identity}
\[
\boxed{\p A+[A,\w]=0}
\]
One can easily check that further differentiation does not add new conditions.

\section{Covariant derivation}

Note that
\[
\p_\Delta f
=[f,\Delta]
=[f,\Delta_0+\omega]
=[f,\Delta_0]+[f,\omega]=\p f+[f,\omega].
\]
One can say that $\nabla:=\p_\Delta$ is a \emph{covariant derivation}.
The Bianci identity reads
\[
\boxed{\nabla A=\p A+[A,\omega]=0}
\]
Also note that
\[
\boxed{\nabla^{2}f=[f,A]}
\]
So the condition $\nabla^{2}=0$ does not imply $A=0$. Instead,
$\nabla^{2}=0$ implies that $A$ lies in the \emph{center} of
$\Com^{-}C$. In particular,
\[
\nabla^{2}=0\quad\Longrightarrow
\quad\p A=0
\quad\Longrightarrow\quad
A\in\Ker\p.
\]
Note that $\Delta$ may nevertheless remain \emph{non-associative}.

\section{Discussion: operadic gravity}

Thus the differential geometrical notion of curvature can be easily
adjusted for deformations in a pre-operad.
Rather than to speak about \emph{algebraic} deformation theory, one may speak
about the \emph{geometrical} one.
Geometry performs the pioneering role in creating of
the exact scientific world picture.
One may ask that how far one can proceed with geometrical notions
in operad theoretical deformation theory.
In particular, this question may be adjusted for physics as well.

In General Relativity, gravity is a fundamental interaction associated with
the space-time curvature (associator \cite{Ki,Ak}). \emph{Operadic}
curvature may be used for representing gravity in a form suitable for
\emph{deformation} quantization. Geodesic multiplication \cite{Ki,Ak} can
here been seen as a prospective model.

One may also follow the Maxwell (gauge field) equations.
It is well-known that the \emph{first} pair of the Maxwell equations
can be represented as the Bianchi identity.
To introduce the \emph{second} pair, one must define a \emph{dualization} $^{\dag}$
and an \emph{operad current} $J$. Then the (gauge field) Maxwell-like equations read
\[
\boxed{ \nabla A=\p A+[A,\w]=0,
\qquad \nabla A^{\dag}=\p
A^{\dag}+[A^{\dag},\w]=J^{\dag} }
\]
In this approach, one must study the physical equations in \emph{non-associative}
deformation complexes.

\section*{Acknowledgements}

I would like to thank  P.~Kuusk, J.~L\~{o}hmus and J.~Stasheff
for helpful comments.
Research was supported in part by the ESF grant 3654.


\begin{thebibliography}{99}
\itemsep-4pt

\bibitem{NeSa1}
\newblock{A.~I.~Nesterov and L.~V.~Sabinin,}
\newblock{Nonassociative geometry: Towards discrete structure of spacetime.}
\newblock{Phys. Rev. {\bf D62} (2000), 081501; hep-th/0010159.}

\bibitem{NeSa2}
\newblock{A.~I.~Nesterov and L.~V.~Sabinin,}
\newblock{Non-associative geometry and discrete structure of spacetime.}
\newblock{Comment. Math. Univ. Carolinae, {\bf 41} (2000), 347-357; hep-th/0003238.}

\bibitem{Ki}
\newblock{M.~Kikkawa,}
\newblock{On local loops in affine manifolds.}
\newblock{J. Hiroshima Univ. Ser. A-1. Math. {\bf28} (1964), 199-207.}

\bibitem{Ak}
\newblock{M.~Akivis,}
\newblock{Geodesic loops and local triple systems in a space with an affine connection.}
\newblock{Sibirski Math. J. {\bf 19} (1978), 243-253 (in Russian).}

\bibitem{Ger}
\newblock{M.~Gerstenhaber,}
\newblock{On cohomology structure of an associative ring.}
\newblock{Ann. of Math. {\bf 78} (1963), 267-288.}

\end{thebibliography}
\end{document}